\documentclass[preprint,aps,showpacs,nofootinbib,preprintnumbers,amsmath,amssymb]{revtex4-1}
\usepackage{}
\usepackage{epsfig}
\usepackage{subfigure}
\usepackage{dcolumn}% Align table columns on decimal point
\usepackage{bm}% bold math
\usepackage[usenames ,dvipsnames]{xcolor}
\usepackage{slashed}
\usepackage{graphicx,color}
\usepackage{centernot}

\begin{document}
\title{Pentaquarks from intrinsic charms in $\Lambda_b$ decays}

\author{Y.K. Hsiao and C.Q. Geng}
\affiliation{
Chongqing University of Posts \& Telecommunications, Chongqing, 400065, China\\
Physics Division, National Center for Theoretical Sciences, Hsinchu, Taiwan 300\\
Department of Physics, National Tsing Hua University, Hsinchu, Taiwan 300
}
\date{\today}

\begin{abstract}
We study the three-body $\Lambda_b$ decays of  $\Lambda_b\to J/\psi pM$ with $M=K^-$ and $\pi^-$.
The two new states ${\cal P}_{c1}\equiv {\cal P}_c(4380)^+$ and ${\cal P}_{c2}\equiv {\cal P}_c(4450)^+$
observed recently as the resonances in the $J/\psi p$ invariant mass spectrum of $\Lambda_b\to J/\psi p K^-$
can be identified to consist of five quarks, $uudc\bar c$, being consistent with the existence of the pentaquark states. 
We argue that, in the doubly charmful $\Lambda_b$ decays of $\Lambda_b\to J/\psi pK^-$ through $b\to c\bar c s$,
apart from  those through the non-resonant $\Lambda_b\to pK^-$ and  resonant $\Lambda_b\to \Lambda^*\to pK^-$ 
transitions, the third contribution with the non-factorizable effects is not the dominant part for the resonant 
$\Lambda_b\to K^-{\cal P}_{c1,c2}, {\cal P}_{c1,c2}\to J/\psi p$ processes, such that we propose that the 
${\cal P}_{c1,c2}$ productions are mainly  from the charmless $\Lambda_b$ decays through $b\to \bar u u s$, 
in which the $c\bar c$ content in ${\cal P}_{c1,c2}$ arises from the intrinsic charms within the $\Lambda_b$ baryon.
We hence predict the observables related to the branching ratios and the direct CP violating asymmetries to be
${\cal B}(\Lambda_b\to \pi^-({\cal P}_{c1,c2}\to) J/\psi p)/{\cal B}(\Lambda_b\to K^-({\cal P}_{c1,c2}\to) J/\psi p)=0.58\pm 0.05$,
${\cal A}_{CP}(\Lambda_b\to \pi^-({\cal P}_{c1,c2}\to)J/\psi p)=(-7.4\pm 0.9)\%$, and 
${\cal A}_{CP}(\Lambda_b\to K^-({\cal P}_{c1,c2}\to)J/\psi p)=(+6.3\pm 0.2)\%$, 
which can alleviate the inconsistency between the theoretical expectations 
from the three contributions in the doubly charmful $\Lambda_b$ decays and the observed data.
\end{abstract}
%\pacs{}

\maketitle
\section{introduction}
\begin{figure}[h]
\centering
\includegraphics[width=2.1in]{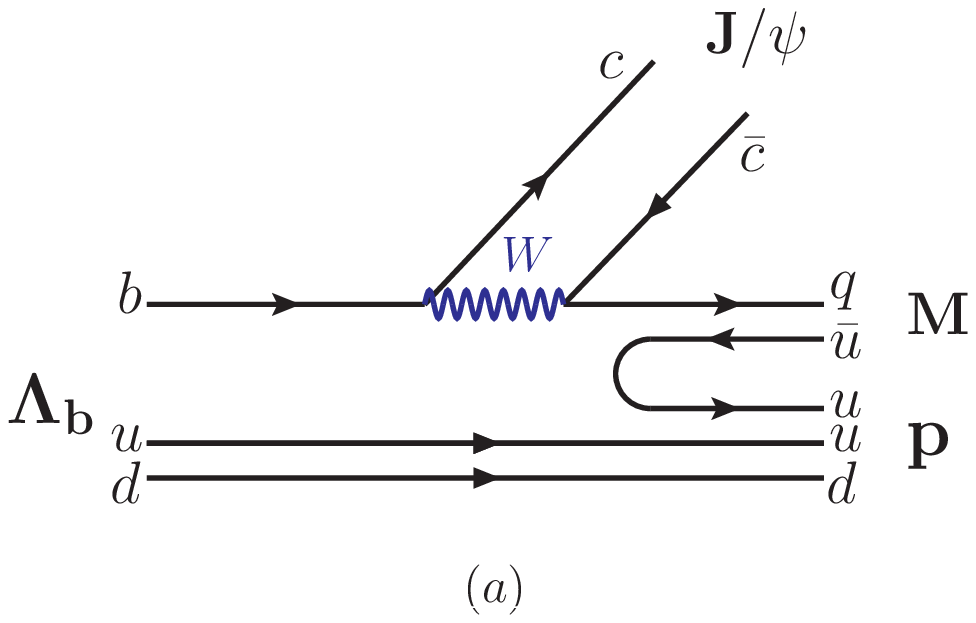}
\includegraphics[width=2.1in]{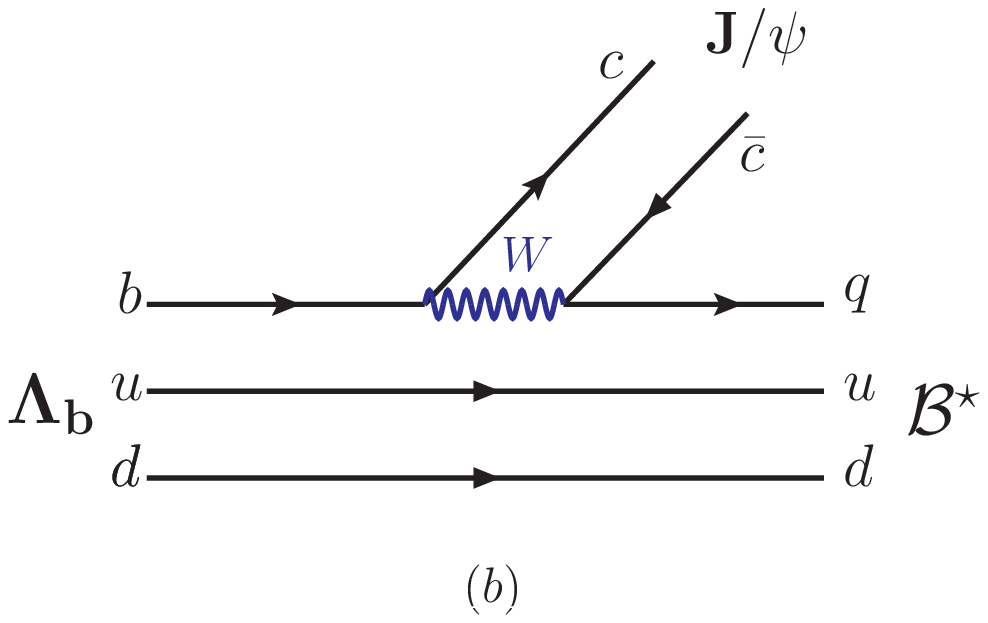}
\includegraphics[width=2.1in]{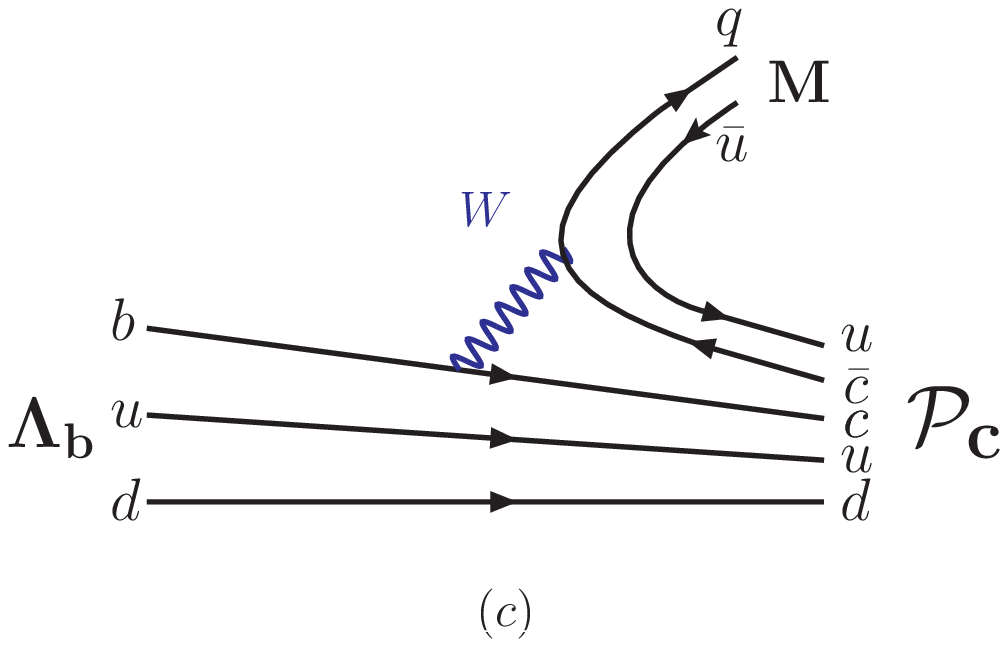}
\caption{Doubly charmful $\Lambda_b$ decays of
$\Lambda_b\to J/\Psi pM$ from
(a) non-resonant $\Lambda_b\to J/\Psi pM$,
(b) resonant $\Lambda_b\to J/\Psi ({\cal B}^\star\to )pM$ with 
${\cal B}^\star\equiv \Lambda^* \to p K^-$ ($N^*\to p\pi^-$) for $q=s(d)$, and 
(c) resonant $\Lambda_b\to M  ({\cal P}_c\to) J/\psi p$ contributions, respectively.}\label{LbtoJpM1}
\end{figure}

According to the recent observations of the three-body $b$-baryon decays of 
$\Lambda_b\to J/\psi pM$ with $M=K^-$ and $\pi^-$~\cite{LHCb1,Pc_LHCb,Aaij:2015fea}, 
apart from the non-resonant $\Lambda_b\to J/\psi p M$ and  resonant
$\Lambda_b\to J/\psi {\cal B}^*,{\cal B}^*\to pM$ (${\cal B}^*$=$\Lambda^*$ ($N^*$) for  $M=K^-$ ($\pi^-$))
contributions,
depicted in Figs.~\ref{LbtoJpM1}a and \ref{LbtoJpM1}b, respectively,
there can be another resonant process
in $\Lambda_b\to J/\psi pM$ as shown in Fig.~\ref{LbtoJpM1}c.
The LHCb collaboration has presented the compelling evidence for 
the new resonant states, being consistent with
the existence of the pentaquark states as the five-quark bound states, while
the ${\cal P}_c(4380)^+$ and ${\cal P}_c(4450)^+$ states are observed as the two resonances
in the $J/\psi p$ invariant mass spectrum of $\Lambda_b\to J/\psi K^- p$, 
with the significance for each state to be more than 9 standard deviations,
which can be regarded to be composed of $uudc\bar c$. We note that, in the same principle,
the two new states should also exist in $\Lambda_b\to J/\psi p\pi^-$.
%=======================
%======================
%

The processes of ${\cal P}_c\to J/\psi p$ in Fig.~\ref{LbtoJpM1}c 
are theoretically known to be dominated by the nonfactorizable effects, calculated 
 non-perturbatively with the scatterings of the soft hadrons, such that
the pentaquarks are considered 
 as the molecular states~\cite{Karliner:2015ina,Chen:2015loa,Liu:2015fea}. 
In spite of the non-factorizable diagrams shown in Fig.~\ref{LbtoJpM1}c, which may get enhanced 
when the strong FSI interactions occur near the threshold to explain
the pentaquark productions,
we propose another possibility based on the factorizable effects.
We note that this type of the processes in Fig.~\ref{LbtoJpM1}c
has not been observed in the previous searches of the lighter pentaquarks than ${\cal P}_c$.
For example, 
the resonant $B^+\to \bar p \Theta(1710)^{++},\Theta(1710)^{++}\to pK^+$ decay
is measured to be 
${\cal B}(B^+\to \bar p \Theta(1710)^{++},\Theta(1710)^{++}\to pK^+)<9.1\times 10^{-8}$
with the upper bound about $60-70$ times smaller than 
the observed branching ratio  of $B^+\to p\bar p K^+$~\cite{Aubert:2005gw,Wang:2005fc},
%PRD72, 051101;PLB617, 141.
while ${\cal B}(B^0\to \bar p \Theta(1540)^+,\Theta(1540)^+\to pK^0_s)<5\times 10^{-8}$
is measured with the upper bound around $50$ times smaller than 
${\cal B}(B^0\to p\bar p K^0_s)$~\cite{Wang:2005fc,Aubert:2007qea}.
%PRD76, 092004;PLB617, 141.
Similar to the charmless $B\to p\bar p K$ decays, 
 the upper bound on  ${\cal B}(B^0\to \Theta_c\bar p\pi^+,\Theta_c\to D^{(*)-}p)$
is expected to be about $30-40$ times smaller than 
${\cal B}(B^0\to p\bar p\pi^+ D^{(*)-})$~\cite{Aubert:2006qx}.
%PRD74, 051101.
In contrast, since the resonant $\Lambda_b\to K^-{\cal P}_c, {\cal P}_c\to J/\psi p$ decays
can contribute to the branching ratio as much as 10\%, 
this leads to the question that if there can be other processes, which are responsible for 
the resonant ${\cal P}_c\to J/\psi P$ decays,
other than the ones in Fig.~\ref{LbtoJpM1}c.

%=======================
\begin{figure}[t!]
\centering
\includegraphics[width=2.2in]{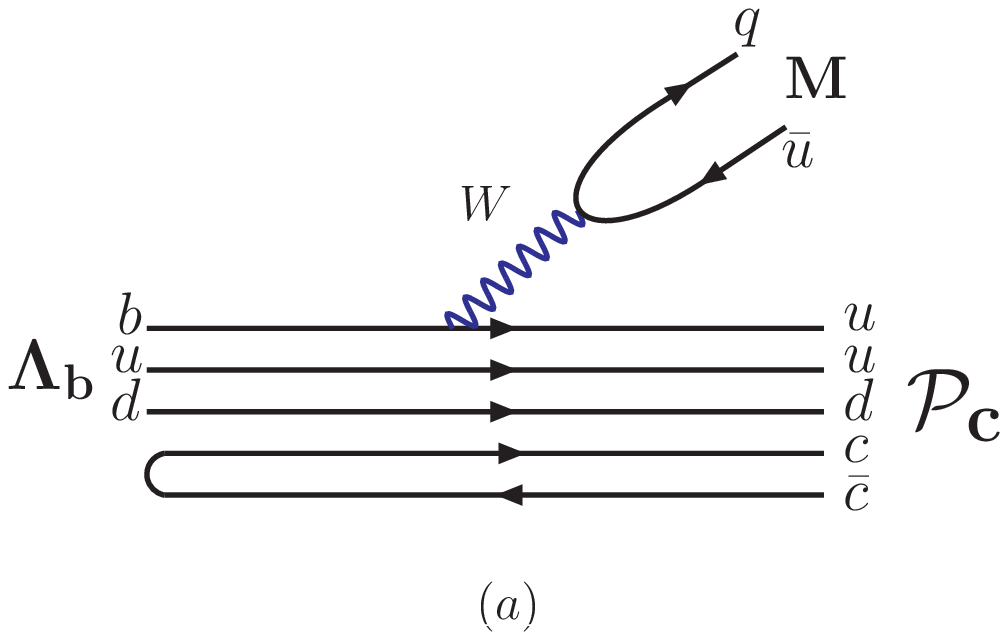}
\includegraphics[width=2.2in]{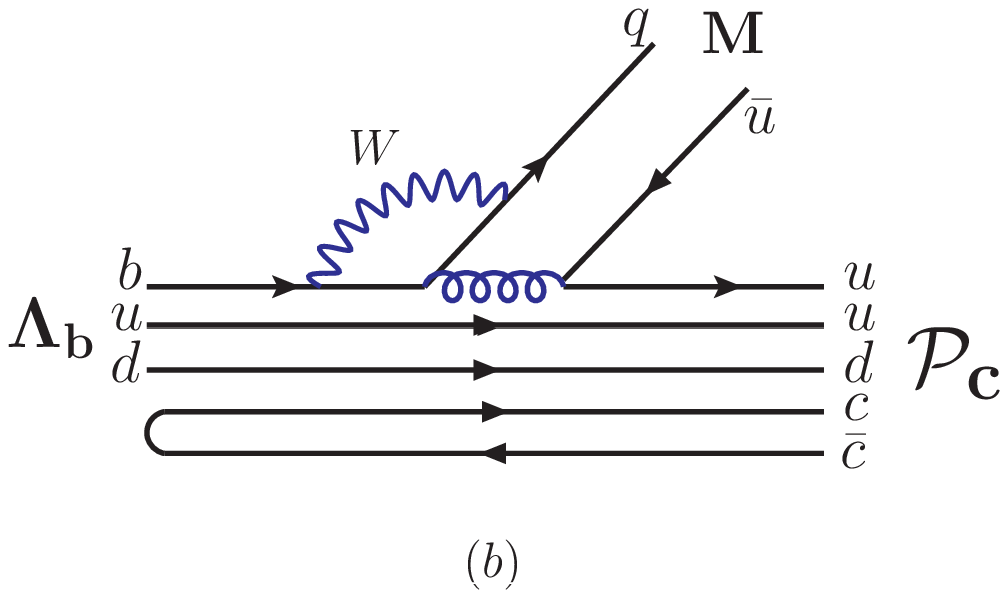}
\caption{The new contributions to the resonant
$\Lambda_b\to M({\cal P}_c\to) J/\Psi p$ 
from the factorizable charmless $\Lambda_b$ decays,
where (a) and (b) are known as the tree and penguin contributions, respectively,
while the $c\bar c$ contents are coming from 
the intrinsic charm within the $\Lambda_b$ baryon.}\label{LbtoJpM2}
\end{figure}
%======================

In Ref.~\cite{LHCb1}, the LHCb collaboration has given
\begin{eqnarray}\label{2relations}
{\cal R}_{\pi K}&\equiv& \frac{{\cal B}(\Lambda_b\to J/\psi p \pi^-)}{{\cal B}(\Lambda_b\to J/\psi p K^-)}
=0.0824\pm 0.0025\pm 0.0042\,,\nonumber\\
\Delta {\cal A}_{CP}&\equiv&{\cal A}_{CP}(\Lambda_b\to J/\psi p \pi^-)-{\cal A}_{CP}(\Lambda_b\to J/\psi p K^-)
=(5.7\pm 2.4\pm 1.2)\%\,,
\end{eqnarray}
where the first and second errors are from the statistical and systematic uncertainties, respectively.
The data in Eq.~(\ref{2relations})
indicate some new $\Lambda_b\to M{\cal P}_c, {\cal P}_c\to J/\psi p$ processes
beyond the non-factorizable ones in Fig.~\ref{LbtoJpM1}c with reasons as follows.

First, we note that ${\cal B}(\Lambda_b\to p \pi^-)/{\cal B}(\Lambda_b\to p K^-)=0.84\pm 0.09$ 
has been theoretically reproduced in Ref.~\cite{Hsiao:2014mua}, 
both ${\cal B}(\Lambda_b\to D^0 p K^-)/{\cal B}(\Lambda_b\to D^0 p \pi^-)
=0.073\pm 0.008^{+0.005}_{-0.006}$
and ${\cal B}(\Lambda_b\to \Lambda_c^+ K^-)/{\cal B}(\Lambda_b\to \Lambda_c^+ \pi^-)
=0.0731\pm 0.0016\pm 0.0016$ can be understood by
the relation of $(V_{us}/V_{ud})^2 (f_K/f_\pi)^2\simeq 0.075$~\cite{Aaij:2013pka}, 
and ${\cal B}(\Lambda_b\to \Lambda_c^+ D^-)/{\cal B}(\Lambda_b\to \Lambda_c^+D_s^-)
=0.042\pm 0.003\pm 0.003$ is not far from 
the relation of $(V_{cd}/V_{cs})^2 (f_D/f_{D_s})^2\simeq 0.035$~\cite{Aaij:2014pha}.
However, ${\cal R}_{\pi K}\simeq 0.08$ in Eq.~(\ref{2relations}) apparently deviates from 
$(V_{cd}/V_{cs})^2\simeq 0.05$ given by
the doubly charmful $\Lambda_b$ decays in Fig.~\ref{LbtoJpM1}. To explain this difference,
some new thinking is needed. 

Second, 
$\Delta {\cal A}_{CP}\sim 5.7\%$ in Eq.~(\ref{2relations}) 
with the significance of 2.2$\sigma$ suggests that
a new contribution must proceed with  $V_{ub}$ to  provide the weak CP phase,
otherwise  $\Delta{\cal A}_{CP}=0$ as the case in the doubly
charmful $\Lambda_b$ decays in Fig.~\ref{LbtoJpM1}, 
in which such a phase is vanishingly small.

We hence propose that the resonant 
$\Lambda_b\to M{\cal P}_c, {\cal P}_c\to J/\psi p$ processes can be the new 
contributions to the charmless $\Lambda_b$ decays as depicted in Fig.~\ref{LbtoJpM2},
where the  $c\bar c$ content  comes from the intrinsic charm (IC) in the $\Lambda_b$ baryon.
In the followings, we will assume that these new processes in Fig.~\ref{LbtoJpM2} are the dominant ones for
$\Lambda_b\to M{\cal P}_c, {\cal P}_c\to J/\psi p$.

It is not  surprising that the $\Lambda_b$ baryon contains the ICs,
which are presented in the Fock state decomposition~\cite{IC1,IC2} as
$|\Lambda_b\rangle=\Psi_{bud}|bud\rangle+\Psi_{budc\bar c}|budc\bar c\rangle+\cdot\cdot\cdot$.
In fact, the existence of the IC  was first suggested in the proton  to explain 
the large $D^+$ and $\Lambda_c^+$ productions
at large energies in the proton-proton scattering~\cite{IC1,IC2}.
In addition, as a possible solution to the so-called $\rho$-$\pi$ puzzle~\cite{Brodsky:1997fj},
the IC in $\rho$ for $J/\psi \to \rho^+\pi^-$ allows a strong decay not through
the $J/\psi$ annihilation suppressed by the OKubo-Zweig-Iizuka (OZI) rule.
For a heavier hadron, since the gluon fluctuation,
such as $gg\to c\bar c$, can easily occur without costing a large energy~\cite{Chang:2001iy},
it is expected that~\cite{Brodsky:2001yt} the IC component in $\Lambda_b$ ($m_{\Lambda_b}>m_B>m_p$)
can be larger than 
the proton and the $B$ mesons, estimated to be 1\% and 4\%, respectively.
Consequently, in the $\Lambda_b\to p$ transition, 
%in spite of the IC content in the proton~\cite{Mikhasenko:2012km}, 
we only consider the ICs in $\Lambda_b$ since the heavier baryon would
contribute a larger $c\bar c$ production.
Note that, to distinguish the IC in the proton
from that in $\Lambda_b$, 
the $J/\psi$ photoproduction can be useful~\cite{Wang:2015jsa,Karliner:2015voa,Kubarovsky:2015aaa},
which is in accordance with Ref.~~\cite{Mikhasenko:2012km}. 
While the study of the ICs in the $B$ decays has been done extensively 
in the literature~\cite{Chang:2001iy,Brodsky:2001yt,Gabbiani:2002ti,Breidenbach:2008ua}, 
it is not well examined in $\Lambda_b$, which should be a suitable scenario.

In this paper, since we propose that the two new resonant ${\cal P}_c$ states, $i.e.$ the pentaquark states, 
in the $m_{J/\psi p}$ spectrum of the $\Lambda_b\to J/\psi p K^-$ decay can be traced back to
the charmless $\Lambda_b$ decays from $b\to u\bar u s$, while the $c\bar c$ content in $J/\psi$
 is from the IC in the $\Lambda_b$ baryon,
we will study  the branching ratios and the direct CP violating asymmetries,
and check if our results will be able to understand the inconsistency between 
the theoretical estimations in the doubly charmful $\Lambda_b$ decays and 
the observed data in Eq.~(\ref{2relations}).

\section{formalism}
In terms of the effective Hamiltonian for $b\to c\bar c q$ at the quark level,
the amplitude of $\Lambda_b\to J/\psi p M$ from Figs.~\ref{LbtoJpM1}a and \ref{LbtoJpM1}b is given by
\begin{eqnarray}\label{amp1}
{\cal A}_{c\bar c q}(\Lambda_b\to J/\psi p M)=\frac{G_F}{\sqrt 2} V_{cb}V_{cq}^* a_2\,
\langle J/\psi|\bar c\gamma^\mu(1- \gamma_5) c|0\rangle
\langle p M|\bar q\gamma_\mu(1-\gamma_5) b|\Lambda_b\rangle\,,
\end{eqnarray}
where $G_F$ is the Fermi constant, $V$ stands for the CKM mixing matrix,
$q=s(d)$ corresponds to $M=K^-(\pi^-)$, 
$\langle p K|\bar s\gamma_\mu(1-\gamma_5) b|\Lambda_b\rangle$ 
contains the contributions from the non-resonant $\Lambda_b\to p M$ and 
 resonant $\Lambda_b\to {\cal B}^*\to p M$ transitions, and
$\langle J/\psi|\bar c\gamma^\mu(1- \gamma_5) c|0\rangle=m_{J/\psi}f_{J/\psi}\varepsilon^{\mu *}$
with $m_{J/\psi}$, $f_{J/\psi}$ and $\varepsilon^{\mu *}$ being the mass, the decay constant
and the polarization of $J/\psi$, respectively. 
Subsequently, the matrix elements of 
the combined $\Lambda_b\to pM$ transition can be parameterized as
\begin{eqnarray}
\langle p M|\bar q\gamma_\mu(1-\gamma_5) b|\Lambda_b\rangle\simeq
F_M e^{i\delta_1} \bar u_p\gamma_\mu(1-\gamma_5)u_{\Lambda_b}\,,
\end{eqnarray}
where $\delta_1$ is the strong phase from the on-shell resonant ${\cal B}^*\to pM$ decay
and $F_M$ is the parameter with
$F_\pi/F_K\simeq (f_\pi/f_K)$ 
representing the flavor $SU(3)$ symmetry breaking.
As a result, 
 we rewrite the amplitude in Eq.~(\ref{amp1}) as
\begin{eqnarray}\label{amp2}
{\cal A}_{c\bar c q}(\Lambda_b\to J/\psi p M)=
\frac{G_F}{\sqrt 2} V_{cb}V_{cq}^* a_2\,
m_{J/\psi}f_{J/\psi} F_M e^{i\delta_1} u_p\centernot{\varepsilon}(1-\gamma_5)u_{\Lambda_b}\,,
\end{eqnarray}
with $\centernot{\varepsilon}=\varepsilon^{\mu *} \cdot\gamma_\mu$.
 From Fig.~\ref{LbtoJpM2}, which depicts the charmless $\Lambda_b$ decays of 
$\Lambda_b\to {\cal P}_c M, {\cal P}_c\to J/\psi M$ decays, 
with $c\bar c$ in ${\cal P}_c$ coming from the IC in $\Lambda_b$,
the amplitudes of $\Lambda_b\to M {\cal P}_c,{\cal P}_c\to J/\psi p$ can be derived as~\cite{ali}
\begin{eqnarray}\label{eq1}
{\cal A}_{{\cal P}_c}
%\equiv{\cal A}(\Lambda_b\to M {\cal P}_c,{\cal P}_c\to J/\psi p)
=i\frac{G_F}{\sqrt 2}m_b f_M\bigg[
\alpha_{M}\langle J/\psi p|\bar u b|\Lambda_b\rangle+
\beta_{M}\langle J/\psi p|\bar u\gamma_5 b|\Lambda_b\rangle\bigg]\,,
\end{eqnarray}
where $f_M$ is the meson decay constant,  defined by
$\langle M|\bar q_1\gamma_\mu \gamma_5 q_2|0\rangle$ $=-if_M q_\mu~$ with
the four-momentum $q_\mu$.
The constant $\alpha_{M}$ ($\beta_M$) in Eq.~(\ref{eq1}) is related to
the (pseudo)scalar quark current, given by
\begin{eqnarray}\label{eq2}
\alpha_{M}(\beta_{M})&=& V_{ub}V_{uq}^*a_1-V_{tb}V_{tq}^*(a_4\pm r_M a_6)\;,
\end{eqnarray}
where $r_M\equiv {2 m_M^2}/[m_b (m_q+m_u)]$
%$V_{ij}$ are the CKM matrix elements,  $q=s(d)$ for $M=K^-(\pi^-)$,
and  $a_i\equiv c^{eff}_i+c^{eff}_{i\pm1}/N_c^{(eff)}$ for $i=$odd (even)
are composed of the effective Wilson coefficients $c_i^{eff}$ defined in Ref.~\cite{ali}.
In Eq.~(\ref{eq1}), the matrix elements for the resonant $\Lambda_b\to {\cal P}_c, {\cal P}_c\to J/\psi p$ 
transition can be given as
\begin{eqnarray}\label{resonantF}
\langle J/\psi p|\bar u(\gamma_5)b|\Lambda_b\rangle=\langle J/\psi p|{\cal P}_c \rangle
%\frac{i}{(t-m_{{\cal P}_c}^2)+im_{{\cal P}_c}\Gamma_{{\cal P}_c}}
{\cal R}_{{\cal P}_c}\langle {\cal P}_c|\bar u(\gamma_5)b|\Lambda_b\rangle\,,
\end{eqnarray}
%with $t=q^2$, 
where the Breit-Wigner factor ${\cal R}_{{\cal P}_c}$ for ${\cal P}_c$ is described as
an intermediate state, given by
\begin{eqnarray}
{\cal R}_{{\cal P}_c}=\frac{i}{(t-m_{{\cal P}_c}^2)+im_{{\cal P}_c}\Gamma_{{\cal P}_c}}\,,
\end{eqnarray}
with $m_{{\cal P}_c}$ and $\Gamma_{{\cal P}_c}$
the mass and the decay width for the ${\cal P}_c$ state, respectively.
Despite the fact that there is no sufficient information for the detailed parameterization of 
$\langle J/\psi p|{\cal P}_c \rangle\langle {\cal P}_c|\bar u(\gamma_5)b|\Lambda_b\rangle$, 
the matrix elements of $\langle J/\psi p|\bar u(\gamma_5)b|\Lambda_b\rangle$ in Eq.~(\ref{resonantF})
can still be reduced as
\begin{eqnarray}
\langle J/\psi p|\bar ub|\Lambda_b\rangle=
{\cal R}_{{\cal P}_c}(\varepsilon \cdot q) F_S\bar u_p u_{\Lambda_b}\,,
\langle J/\psi p|\bar u\gamma_5 b|\Lambda_b\rangle=
{\cal R}_{{\cal P}_c}(\varepsilon \cdot q)F_P\bar u_p \gamma_5u_{\Lambda_b}\,.
\end{eqnarray}
This is due to the fact that, 
after the summations of the intermediate ${\cal P}_c$ spins with spin=$3/2$ or $5/2$,
all Lorentz indices are in fact coupled to be a scalar quantity,
which can be parameterized as $F_S$ and $F_P$. 
In general,
$F_{S,P}$ are  momentum dependent, but they can be taken as nearly constants
around the threshold area of $t\simeq m_{{\cal P}_c}^2$, at which the threshold effect dominates 
the decay branching ratio.
Besides, we take $F_S=F_P\equiv F_{{\cal P}_c}$ 
as a consequence of the $\Lambda_b$ transition~\cite{Hsiao:2014mua}.
We hence obtain
${\cal A}_{{\cal P}_c}
\simeq i\frac{G_F}{\sqrt 2}m_b f_M {\cal R}_{{\cal P}_c}F_{{\cal P}_c}
\bar u_p (\alpha_{M}+\beta_{M}\gamma_5)u_{\Lambda_b}$,
such that the total amplitude for the two resonant ${\cal P}_c$ states is in the form of
\begin{eqnarray}
{\cal A}(\Lambda_b\to M({\cal P}_{c1},{\cal P}_{c2}\to)J/\psi p)
%={\cal A}_{{\cal P}_c(4380)^+}+{\cal A}_{{\cal P}_c(4450)^+}
={\cal A}_{{\cal P}_{c1}}+{\cal A}_{{\cal P}_{c2}}
\simeq i\frac{G_F}{\sqrt 2}m_b f_M F_2e^{i\delta_2}\,\bar u_p (\alpha_{M}+\beta_{M}\gamma_5)u_{\Lambda_b}\,,~
\label{eq10}
\end{eqnarray}
with $F_2 e^{i\delta_2}={\cal R}_{{\cal P}_{c1}}F_{{\cal P}_{c1}}+
{\cal R}_{{\cal P}_{c2}}F_{{\cal P}_{c2}}$,
where $\delta_2$ is the strong phase from the on-shell ${\cal P}_c\to J/\psi p$ decays, and
${\cal P}_{c1}$ and ${\cal P}_{c2}$ denote ${\cal P}_c(4380)^+$ and ${\cal P}_c(4380)^+$,
respectively.
Note that ${\cal P}_{c1,c2}$ have been observed to have the masses and the decay widths 
as $(m,\Gamma)=(4380\pm 8\pm 29,\,205\pm 18\pm 86)$ MeV and 
$(4449.8\pm 1.7\pm 2.5,\,39\pm 5\pm 19)$ MeV, respectively,
while their quantum numbers for $J^P$ can be $(3/2^-,\,5/2^+)$ or $(3/2^+,\,5/2^-)$.
However, the information of ${\cal P}_{c1,c2}$ can be cast into 
the to-be-determined parameters $F_2 e^{i\delta_2}$, without losing generality.

\section{Numerical analysis and Discussions}
For the numerical analysis,
the theoretical inputs of the meson decay constants and
Wolfenstein parameters in the CKM matrix are taken
as~\cite{pdg,Becirevic:2013bsa} 
\begin{eqnarray}
&&(f_{J/\psi},f_\pi,f_K)=(418\pm 9,\,130.4\pm 0.2,\,156.2\pm 0.7)\,\text{MeV}\,,\nonumber\\
&&(\lambda,\,A,\,\rho,\,\eta)=
(0.225,\,0.814,\,0.120\pm 0.022,\,0.362\pm 0.013)\,,
\end{eqnarray}
while the parameters $a_{1,4,6}$ can be adopted from Refs.~\cite{GengHsiao,Hsiao:2014mua},
%and $a_2$ is adopted to be 
along with $a_2=0.2$~\cite{Chen:2008sw}.
The data  in the fitting are given in Table~\ref{fitdata}.
\begin{table}[b]%[htb]
\caption{The experimental inputs for the fitting, 
where the numbers are taken from \cite{LHCb1,Pc_LHCb}.}
\label{fitdata}
\begin{tabular}{|c|c|c|}
\hline
&data&fitting results\\\hline
${\cal R}_{\pi K}$&
$(8.24\pm 0.49)\%$&
$(8.38\pm 0.77)\%$\\
$\Delta{\cal A}_{CP}$&
$(5.7\pm 2.7)\%$&
$(2.9\pm 1.4)\%$\\
$10^4 {\cal B}(\Lambda_b\to K^- J/\psi p)$&$3.04\pm 0.55$&$3.21\pm 0.44$\\
%
%$\frac{{\cal B}(\Lambda_b\to K^-({\cal P}_{c2}\to) J/\psi p)}{{\cal B}(\Lambda_b\to K^- J/\psi p)}$&
%$(4.1\pm 1.2)\%$&\\
%
$10^6 {\cal B}(\Lambda_b\to K^-({\cal P}_{c1}\to) J/\psi p)$&
$25.6\pm 13.8$&$10.3\pm 3.9$\\
$10^6 {\cal B}(\Lambda_b\to K^- ({\cal P}_{c2}\to) J/\psi p)$&
$12.5\pm 4.2$&
$10.9\pm 2.7$\\
\hline
\end{tabular}
\end{table}
As a result, we obtain
\begin{eqnarray}
F_K=2.8\pm 0.2\,,\;
F_{{\cal P}_{c1}}=19.6\pm 3.1\,,\;
F_{{\cal P}_{c2}}=5.5\pm 1.0\,,
\delta_1=(54.8\pm 31.9)^\circ\,,
\end{eqnarray}
%{2.784, 19.576, 5.536, 54.752};
%{0.221, 3.116, 0.952, 31.851};
with the fitted numbers in column 2 of Table~\ref{fitdata} 
to be consistent with the data.
First, 
for the three-body $\Lambda_b$ decays only from
the resonant $\Lambda_b\to M{\cal P}_c, {\cal P}_c\to J/\psi p$ contributions in Fig.~\ref{LbtoJpM2},
we obtain 
\begin{eqnarray}
\frac{{\cal B}(\Lambda_b\to \pi^-({\cal P}_{c1,c2}\to) J/\psi p)}
{{\cal B}(\Lambda_b\to K^-({\cal P}_{c1,c2}\to) J/\psi p)}=0.58\pm 0.05\,,
\end{eqnarray}
where  the parameters $F_2 e^{i\delta_2}$ in Eq.~(\ref{eq10}) have been canceled by the ratio. 
In the doubly charmful $\Lambda_b$ decays,
since the three contributions are all through $b\to c\bar c q$ at the quark level
(see Fig.~\ref{LbtoJpM1}),
the ratio of ${\cal R}_{\pi K}$ defined in Eq.~(\ref{2relations}) should be
%${\cal R}_{\pi K}=
$(V_{cd}/V_{cs})^2\simeq 0.05$, which is not approved by the data in Eq.~(\ref{2relations}).
However, by adding the contributions from the charmless decays of 
$\Lambda_b\to K^-({\cal P}_c\to)J/\psi p$,
%the data in Ref.~\cite{Pc_LHCb} show
%${\cal B}(\Lambda_b\to K^-({\cal P}_c\to)J/\psi p)$ 
%fitted to be 5\% of the total branching ratio,
the doubly charmful ${\cal B}(\Lambda_b\to J/\psi p\pi^-)$ 
is getting close to the charmless ${\cal B}(\Lambda_b\to J/\psi({\cal P}_c\to)p\pi^-)$, 
so that the value of ${\cal R}_{\pi K}$ is able to increase from
$0.05$ to a larger one  to meet the data in Eq.~(\ref{2relations}),
as the fitted result of $(8.38\pm 0.77)\%$ in Table~\ref{fitdata}.

Second, the direct CP violating asymmetries from
the resonant $\Lambda_b\to M{\cal P}_c, {\cal P}_c\to J/\psi p$ parts
are evaluated to be
\begin{eqnarray}\label{preCP}
{\cal A}_{CP}(\Lambda_b\to \pi^-({\cal P}_{c1,c2}\to)J/\psi p)&=&(-7.4\pm 0.9)\%\,,\nonumber\\
{\cal A}_{CP}(\Lambda_b\to K^-({\cal P}_{c1,c2}\to)J/\psi p)&=&(+6.3\pm 0.2)\%\,.
\end{eqnarray}
However, since the measurement by the LHCb in Ref.~\cite{Pc_LHCb} has  suggested that 
the doubly charmful 
  $\Lambda_b\to J/\psi p K^-$ mode
dominates the corresponding decay,
it leaves %a 
little room for the interference effects with the charmless ones of 
$\Lambda_b\to K^-({\cal P}_{c1},{\cal P}_{c2}\to)J/\psi$ that provide the weak CP phase, 
of which ${\cal A}^{\text{raw}}_{CP}(\Lambda_b\to J/\psi p K^-)=(1.1\pm 0.9)\%$ 
from the LHCb~\cite{LHCb1} agrees with the  fitted result of 
${\cal A}_{CP}(\Lambda_b\to J/\psi p K^-)=(-0.22\pm 0.16)\%$.
%Moreover,
%the data of ${\cal A}^{\text{raw}}_{CP}(\Lambda_b\to J/\psi p K^-)=(1.1\pm 0.9)\%$ from the LHCb~\cite{LHCb1}
%is consistent with zero, whereas the value for $K^-$ in Eq.~(\ref{preCP}) arises from
%the charmless $\Lambda_b\to K^-({\cal P}_{c1,c1}\to)J/\psi p$ only.
%Nonetheless, 
Note that $\Delta {\cal A}_{CP}=0$ from $b\to c\bar c q$
% the doubly charmful $\Lambda_b$ decays
to be different from the data of $\Delta {\cal A}_{CP}=5.7\%$ in Eq.~(\ref{2relations})
%${\cal A}^{\text {raw}}_{CP}(\Lambda_b\to J/\psi p \pi^-)=(7.9\pm 2.2)\%$~\cite{LHCb1}
requires the interference between the two compatible $\Lambda_b\to \pi^-({\cal P}_{c1,c2}\to)J/\psi p$ and 
$\Lambda_b\to J/\psi (N^*(1440),N^*(1520)\to)p\pi^-$ channels.
%which can clearly change the sign and size of that 
%
%in Eq.~(\ref{preCP}) for the $\pi$ mode.
%This is the reason why $\Delta{\cal A}_{CP}=(5.7\pm 2.7)\%$ in Table~\ref{fitdata}
%Eq.~(\ref{2relations})
%can be different from $\Delta{\cal A}_{CP}\simeq -14\%$ from Eq.~(\ref{preCP}).
%In fact, 
It is found that
the contributions from $b\to c\bar c d$ with the strong phase $\delta_1=54.8^\circ$
and the contributions from $b\to u\bar u d$ with the weak phase by $V_{ub}$
gives $\Delta{\cal A}_{CP}=(2.9\pm 1.4)\%$, 
which is in good agreement with the data.
Finally, the branching ratio for $\Lambda_b\to J/\psi p \pi^-$ is predicted as
\begin{eqnarray}
{\cal B}(\Lambda_b\to J/\psi p \pi^-)&=&(2.68\pm 0.34)\times 10^{-5}\,,%\nonumber\\
%{\cal B}(\Lambda_b\to J/\psi p K^-)&=&(3.21\pm 0.44)\times 10^{-4}\,,
\end{eqnarray}
which includes the compatible contribution from 
$\Lambda_b\to \pi^-({\cal P}_{c1,c2}\to)J/\psi p$ to agree well with 
${\cal B}(\Lambda_b\to J/\psi p \pi^-)=
(2.51\pm 0.08\pm 0.13^{+0.45}_{-0.35})\times 10^{-5}$ measured by the LHCb~\cite{Aaij:2015fea}, whereas
the contributions only from the doubly charmful $\Lambda_b$ decays give
${\cal B}(\Lambda_b\to J/\psi p \pi^-)=(1.68\pm 0.24)\times 10^{-5}$, which is
around 0.05${\cal B}(\Lambda_b\to J/\psi p K^-)$, borne by the relation of $|V_{cd}/V_{cs}|^2$.

In sum,  the charmless processes of $\Lambda_b\to M({\cal P}_{c1},{\cal P}_{c2}\to)J/\psi$ 
provide us with a possible way to understand the CP asymmetry in Eq.~(\ref{2relations}) due to the origin
of the weak phase from $V_{ub}$. Furthermore, to realize the ratio of
${\cal R}_{\pi K}$ in Eq.~(\ref{2relations}), which  is unable to be explained from $b\to c\bar c q$,
% can be realized, which indicates 
the contributions apart from the non-perturbative processes in Fig.~\ref{LbtoJpM1}c have to be included.

\section{Conclusions}
Since the non-factorizable effects for the doubly charmful $\Lambda_b$ decays
through $b\to c\bar c s$
may not be  suitable to understand the resonant $\Lambda_b\to K^-{\cal P}_c, {\cal P}_c\to J/\psi p$ decays,
while the new ${\cal P}_c$ states observed in $m_{J/\psi p}$ spectrum can be identified 
as the pentaquark states with
five quarks, $uudc\bar c$, 
we have proposed that these resonant 
%$\Lambda_b\to K^-{\cal P}_c, {\cal P}_c\to J/\psi p$ 
processes
could proceed as the charmless $\Lambda_b$ decays through $b\to u\bar us$, while
the $c\bar c$ content in the ${\cal P}_c$ states is from the intrinsic charms in the $\Lambda_b$ baryon.
As a result, we predicted that 
${\cal B}(\Lambda_b\to \pi^-({\cal P}_{c1,c2}\to) J/\psi p)/
{\cal B}(\Lambda_b\to K^-({\cal P}_{c1,c2}\to) J/\psi p)=0.58\pm 0.04$,
${\cal A}_{CP}(\Lambda_b\to \pi^-({\cal P}_{c1,c2}\to)J/\psi p)=(-7.4\pm 0.9)\%$, and 
${\cal A}_{CP}(\Lambda_b\to K^-({\cal P}_{c1,c2}\to)J/\psi p)=(+6.3\pm 0.2)\%$,
which could alleviate the inconsistency between  
the theoretical expectations of $({\cal R}_{\pi K},\Delta A_{CP})=( 0.05,\,0)$
 in the doubly charmful $\Lambda_b$ decays
and the observed data of 
$({\cal R}_{\pi K},\Delta A_{CP})\sim (0.08,5.7\%)$.
%(0.0824\pm 0.0025\pm 0.0042,\,(5.7\pm 2.4\pm 1.2)\%)$.

\section*{ACKNOWLEDGMENTS}
The authors would like to thank Yuan-Ning Gao, and Qiang Zhao for useful discussions.
The work was supported in part by National Center for Theoretical Science, National Science
Council (NSC-101-2112-M-007-006-MY3), MoST (MoST-104-2112-M-007-003-MY3) and National Tsing Hua
University (104N2724E1).

\end{document}